\documentclass[journal,comsoc]{IEEEtran}

\ifCLASSINFOpdf

\else

\fi

\usepackage{mathrsfs}
\usepackage{amsmath}
\usepackage{amsfonts}
\usepackage{amsthm}
\usepackage{amssymb}
\usepackage{graphicx}
\usepackage{subfigure}
\usepackage{indentfirst}
\usepackage{array}
\usepackage{epstopdf}
\usepackage{cite}
\usepackage{enumerate}
\usepackage{bm}
\usepackage{dsfont}
\usepackage[linesnumbered,ruled,vlined]{algorithm2e}
\usepackage{multirow}
\usepackage{verbatim}
\usepackage{stfloats}
\usepackage[table]{xcolor}
\usepackage{makecell}
\usepackage{cancel}

\SetKwComment{Comment}{//}{}
\SetKwBlock{Begin}{Function}{end}

\begin{document}


\title{\huge{NeurJSCC Enabled Semantic Communications:\\Paradigms, Applications, and Potentials}}

\author{Sixian Wang,
        Jincheng Dai, \IEEEmembership{Member, IEEE},
        Xiaoqi Qin, \IEEEmembership{Member, IEEE},
        Kai Niu, \IEEEmembership{Member, IEEE},
        and Ping Zhang, \IEEEmembership{Fellow, IEEE}



\thanks{The authors are with Beijing University of Posts and Telecommunications, Beijing 100876, China. Corresponding authors: Jincheng Dai, Kai Niu. The arXiv preprint version of this article has been given.}

\vspace{-2em}
}

\maketitle

\begin{abstract}
Recent advances in deep learning have led to increased interest in solving high-efficiency end-to-end transmission problems using methods that employ the nonlinear property of neural networks. These techniques, we call \emph{neural joint source-channel coding (NeurJSCC)}, extract latent semantic features of the source signal across space and time, and design corresponding variable-length NeurJSCC approaches to transmit latent features over wireless communication channels. Rapid progress has led to numerous research papers, but a consolidation of the discovered knowledge has not yet emerged. In this article, we gather diverse ideas to categorize the expansive aspects on NeurJSCC as two paradigms, i.e., \emph{explicit} and \emph{implicit} NeurJSCC. We first focus on those two paradigms of NeurJSCC by identifying their common and different components in building end-to-end communication systems. We then focus on typical applications of NeurJSCC to various communication tasks. Our article highlights the improved quality, flexibility, and capability brought by NeurJSCC, and we also point out future directions.
\end{abstract}

\IEEEpeerreviewmaketitle

\section{Introduction of NeurJSCC}\label{section_introduction}

The design goal of end-to-end communication is to reduce the cost of channel bandwidth needed to transmit useful source information. To this end, guided by Shannon's \emph{separation theorem} \cite{cover1999elements}, most traditional systems employ a two-step encoding process to transmit the source data (e.g., audio, speech, image, video, etc.): the source data is first compressed with a source coding algorithm to remove its inherent redundancy, and get an efficient representation of the source data; the compressed bit stream is further encoded with an error correcting code which enables resilient transmission against errors caused by imperfect wireless channels, and modulated as channel-input symbols. However, the separation theorem proves that the two-step source and channel coding approach can only be optimal theoretically given the asymptotic limit of infinite source and channel coding blocks. In practical applications, joint source-channel coding (JSCC) \cite{cover1999elements} is known to outperform the separate approach, but the separate architecture is also attractive thanks to the modularity and simplicity it provides. Classical hand-crafted JSCC has not been easy to compete with advanced separate coding approaches due to the intractable distribution of realistic source data. Emerging \emph{neural}, or \emph{learned} JSCC, is the application of neural networks and related machine learning techniques to achieve JSCC, which however can be end-to-end optimized to capture complex source distribution and presents superior performance.

Neural JSCC (NeurJSCC) enabled end-to-end communication involves feature extraction, feature transmission, feature synthesis of data about objects and scenes across space-time \cite{gunduz2022beyond}. The whole process incorporates the effects of imperfect wireless channels, the goals of downstream tasks, the trade-off between channel bandwidth cost and performance, etc. Within these disciplines, \emph{NeurJSCC} lies at the heart in the emerging end-to-end semantic communication systems, which is responsible for \emph{feature extraction}, \emph{feature coded transmission}, and \emph{feature synthesis} toward human perception or machine tasks \cite{dai2022nonlinear}. Different from traditional JSCC architectures, NeurJSCC seeks for an end-to-end optimization of both compression and error correction functions by leveraging the learning property of neural networks. Moreover, NeurJSCC leverages the nonlinear property of neural networks to provide much higher model capacity for an improved transmission quality. Rapid progress in NeurJSCC has led to numerous papers, but a consolidation of the discovered knowledge has not emerged, making it hard to communicate ideas and train new researchers. Furthermore, there is selective amnesia for old or even concurrent works, requiring new summarization perspectives.

We address above issues by defining NeurJSCC in the light of its basic \emph{signal representation mechanism}. We justify that in end-to-end semantic communication systems, the semantics of information is not necessarily defined as the meaning of data that serves for human understanding, but as the \emph{significance}\footnote{The word \emph{semantics} roots from the Greek ``\emph{sema}'', meaning ``\emph{sign},'' and its related adjective, ``\emph{semantikos}'', meaning ``\emph{significant}.'' It has later evolved to ``meaning'' in the context of languages. Our definition and usage align with the etymology of semantics.} of both data and environment relative to the goal attainment. We justify that NeurJSCC includes both \emph{explicit} and \emph{implicit} paradigms, in which case we provide a shared principle formulation across common techniques, categorizing, describing, and relating their application scenarios. Specifically, we divide our article as two main parts. In part one, we describe several typical paradigms of NeurJSCC that are common and different in terms of their signal representation mechanisms. We identify the commonalities, connections, and trends across NeurJSCC techniques toward different signal modalities and communication tasks. In part two, we present a comprehensive comparison and exemplary applications of NeurJSCC. In summary, NeurJSCC techniques lie at the heart of end-to-end semantic communication systems, this article provides the comprehensive formulation of techniques and discussion of applications to highlight the significance of this emerging area and motivate researchers to make explorations in the future.

\begin{figure*}[t]
	\centering{\includegraphics[scale=0.4]{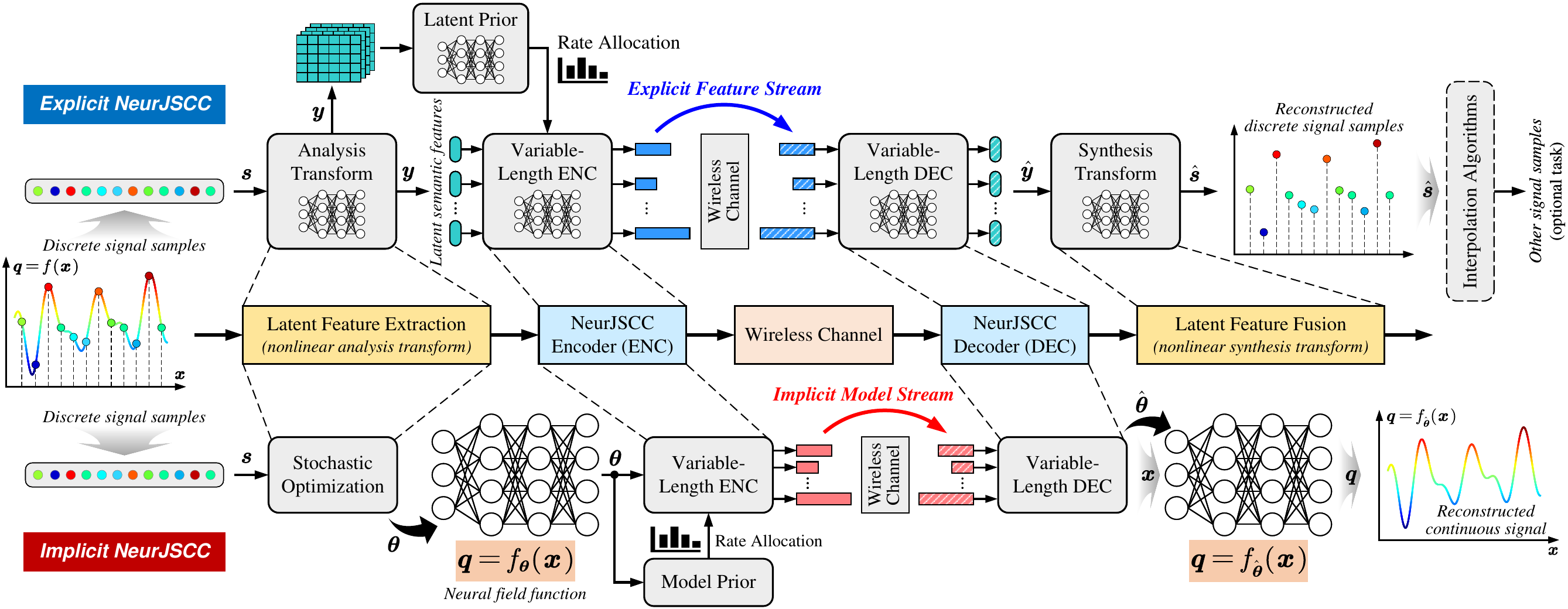}}
	\caption{A system architecture of NeurJSCC, where explicit and implicit NeurJSCC paradigms specify each module as different functions.}\label{Fig1}
	\vspace{-1em}
\end{figure*}

\section{Paradigms of NeurJSCC}

How we represent signals has a tremendous impact on how we solve coded transmission problems in end-to-end communication systems. From the perspective of signal representation mechanisms, we discuss several NeurJSCC techniques.

\subsection{Explicit NeurJSCC}

Both explicit and implicit NeurJSCC for end-to-end data transmission proceed as depicted in Fig. \ref{Fig1}. For explicit NeurJSCC, most existing studies are inspired by the paradigm of auto-encoder \cite{bourtsoulatze2019deep}. Across space-time, we collect source signal samples (data) and feed them into a neural network based \emph{nonlinear analysis transform} to produce latent features, which are indeed an explicit representation of source signal. These latent features are further sent into a \emph{NeurJSCC encoder} to produce the channel-input sequence. During this procedure, a crucial \emph{latent prior}, or \emph{learned entropy model}, is imposed on these latent features, it variationally estimates the entropy distribution on the latent space, which indicates the \emph{significance} (importance) of latent features \cite{dai2022nonlinear}. It guides the NeurJSCC rate/power allocation and transmission resource scheduling to maximize the system coding gain, which becomes the prime gain of NeurJSCC.

In the light of representation learning, explicit NeurJSCC is more suitable for common \emph{structured data} (discrete sampling of a signal over a regular lattice, such as audio/speech, image, and video) since well-developed deep neural networks (such as convolutional neural network (CNN), Transformer, etc.) can efficiently process these signals. Explicit NeurJSCC focuses on mimicking the functionality of traditional discrete signal representation and transmission in a differentiable manner, often using regular discrete grids of learned latent features. Based on the optimized nonlinear transform and NeurJSCC codec modules, the whole amortized model can present good generalization for a given range of data and channel states.

During this process, many problems affect our model ability to transmit the latent features, and reconstruct the source signal with high fidelity or execute the downstream tasks with high accuracy. To better apply explicit NeurJSCC, we identify two classes of useful techniques as follows.

\subsubsection{\underline{Latent Prior Modeling}}

Suppose we wish to reconstruct the signal from received latent features. This problem arises in achieving a lower end-to-end distortion given the constraint of channel bandwidth cost, i.e., we expect a better end-to-end rate-distortion (RD) trade-off. To this end, we need an entropy estimation on the latent features as accurate as possible. One efficient solution is incorporating a scalable hyperprior as the side information, which variationally estimates the parameters of Gaussian distribution on latent space. Apart from these forward adaption (FA) based variational auto-encoder methods \cite{balle2020nonlinear}, we can also introduce the backward adaption (BA) mechanism to derive the contextual entropy model to further capture the dependencies among latent features.


\subsubsection{\underline{Neural Network Architectures}}

Neural network architectures are also important to push the performance of explicit NeurJSCC. Initial works built nonlinear transforms and NeurJSCC codec upon CNNs for images, but they cannot capture the global dependencies among signal samples. Transformer architectures based on attention mechanism can capture long-range dependencies. They can be applied to build components in explicit NeurJSCC, as a result, we obtain more compact latent representations which foster more efficient JSCC.

\subsection{Implicit NeurJSCC}

Implicit NeurJSCC refers to a class of representation techniques to parameterise signals fully or in part using neural networks. The key idea is leveraging neural network parameters $\boldsymbol{\theta}$ to represent the signal using the overfitting property of neural networks \cite{xie2022neural}, after which variable-length NeurJSCC encoding strategy is deigned to transmit model parameters efficiently. In this paradigm, the source feature extraction module in Fig. \ref{Fig1} is performed as an implicit neural representation (INR) function which is trained to map each point in a given domain to the corresponding value of a signal at that point. Structured data can also be represented using INR by mapping the coordinate to the corresponding signal value, where INR is specialized as a coordinate-to-value function. For example, INRs for images learn to map the 2D coordinates of pixels to their corresponding RGB values. Apart from structured signals, INRs are more expert in modeling complex \emph{unstructured signals}, where signal samples rely on unconstrained displacement of points to freely represent arbitrary-shaped 3D objects, etc. In that case, INR is often adopted as a coordinate-to-attribute function. Such a class of INR functions $\boldsymbol{q} = f_{\boldsymbol{\theta}}(\boldsymbol{x})$ with spatial and/or temporal coordinate $\boldsymbol{x}$ as input and physical quantities $\boldsymbol{q}$ as output is named \emph{neural field} \cite{xie2022neural}, which have been successfully applied for modeling 3D shapes and scenes including signed distance functions (SDF) \cite{park2019deepsdf} and neural radiance fields (NeRF) \cite{mildenhall2021nerf}.

Implicit NeurJSCC enjoys several appealing properties:
\begin{itemize}
  \item \emph{Continuous signal representation ability:} Explicit NeurJSCC represents a signal as discrete samples on pixel or voxel grids, where the trade-off is controlled by lattice resolution. However, our world is continuous, based on this observation, INR learns to model complex signal as a function defined in a continuous domain such that the signal information is embedded in the neural network parameters. Accordingly, we can transmit and generate signals at arbitrary resolution we want.

  \item \emph{Instinctive ability to encode a wide range of data modalities:} Explicit NeurJSCC relies on auto-encoders optimized with entropy penalty. For good performance, these algorithms rely heavily on the neural codec architectures tailored for a given data format. Applying a well-learned model to new data modalities requires new designs which are usually challenging. Nevertheless, implicit NeurJSCC bypasses this problem by fitting a neural network mapping coordinates to attributes directly. It builds flexible neural functions compatible to a range of modalities.

  \item \emph{Eliminating concerns on model generalization:} A major concern of existing explicit NeurJSCC works is that an amortized neural codec model cannot generalize well over diverse source/channel samples. However, the encoding process of implicit NeurJSCC works by online overfitting a model to every instance of source signal and channel state, which naturally eliminates the concerns of generalization.
\end{itemize}

To better apply implicit NeurJSCC, we also identify three classes of useful techniques as follows.

\subsubsection{\underline{Periodic Activation Functions}}

As an approximator for low-dimensional-but-complex functions, it is sufficient to realize neural field using coordinate-based multi-layer perceptrons (MLP) rather than other complex neural network architectures adopted in implicit NeurJSCC codec. Key to the successes of coordinate-based MLP is leveraging periodic activation functions as analyzed in \cite{sitzmann2020implicit}. Sinusoidal representation networks (SIREN) have been widely recognized as an effective solution that uses sinusoidal activation function rather than the simple ReLU activation function.

\subsubsection{\underline{Positional Encoding}}

To improve the convergence speed of neural fields, positional encoding can be applied on the input coordinates to map them to a higher dimensional space before feeding into the coordinate-based MLP. Specifically, the Fourier feature (combined with a set of sinusoidal functions) based positional mapping is often employed in INR \cite{tancik2020fourier}.

\subsubsection{\underline{Meta-Learning Techniques}}

Recent advances in meta-learning provide a good idea to accelerate INR signal representation as well as yield better performance. We can use model-agnostic meta-learning (MAML) to find model parameter initializations that can generate INR function with fewer gradient updates so as to speed up implicit NeurJSCC encoding.

We manifest that implicit NeurJSCC is a promising signal representation and transmission paradigm that is quite expert on emerging data such as extremely high resolution images and videos, 3D objects and scenes in XR, etc.

\subsection{Hybrid NeurJSCC}

\begin{figure*}[t]
	\centering{\includegraphics[scale=0.42]{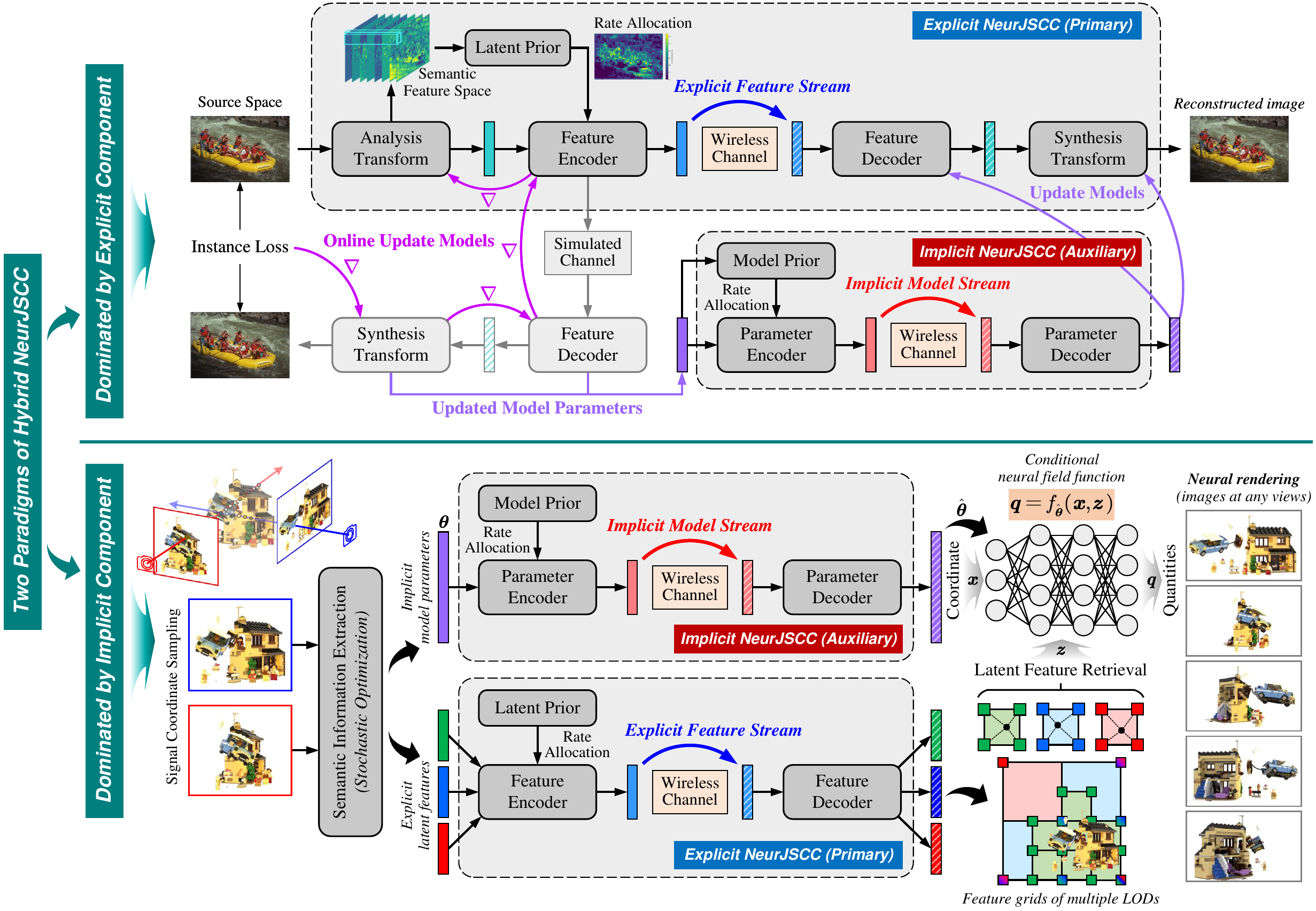}}
	\caption{System architectures of two paradigms of hybrid NeurJSCC transmission.}\label{Fig2}
	\vspace{-1em}
\end{figure*}

To transmit complex signals, limited by the model capacity and signal fitting ability, one may not rely on explicit or implicit NeurJSCC solely. To tackle this, we wish to boost signal reconstruction quality via \emph{prior learning and conditioning}, by this means, \emph{hybrid NeurJSCC} will be more preferred. It combines the advantages of explicit and implicit NeurJSCC together. Specifically, we manifest two types of hybrid NeurJSCC, which are dominated functionally by explicit NeurJSCC and implicit NeurJSCC, respectively.

\subsubsection{\underline{Hybrid NeurJSCC Dominated by Explicit Component}}

Existing optimized explicit NeurJSCC has proven to be quite successful in minimizing the end-to-end expected RD cost over a source dataset and ergodic channel responses, they are yet unlikely to be optimal for every test instance due to the limited model capacity. Existing models only pay attention to an average low RD cost on the training set. For a given instance of source data and channel response, such an amortized codec might not be good at capturing the data latent features and channel state at this instance, resulting in suboptimal transform and coding during the model inference stage. This imperfect optimization and generalization will be especially severe when the testing data distribution or channel response is different from that in the training phase. To tackle this, we can explore a different approach by optimizing NeurJSCC codec and latent representations individually, on a per data sample and channel response basis, during the model inference stage. In other words, we turn to minimize the \emph{instant RD} cost for substantial gains on every data and channel state instance. Essentially, we introduce the overfitted neural-enhancement mechanism to explicit NeurJSCC for profit, which builds adaptive semantic communication (ASC) system.

How to incorporate implicit NeurJSCC as auxiliary components in ASC systems? As illustrated in Fig. \ref{Fig2}, since the NeurJSCC decoder and the nonlinear synthesis transform module are updated online (marked with back-propagated gradient flow $\nabla$ in Fig. \ref{Fig2}), their updated model parameters will also be transmitted over wireless channels. Similar to the explicit NeurJSCC pipeline to transmit data, we also incorporate a \emph{model prior} to estimate the distribution of model parameter updates, accordingly, an efficient variable-length NeurJSCC strategy can be designed to transmit model updates. Compared to the bandwidth cost adopted for transmitting explicit latent features, the bandwidth cost of implicit coding to transmit model updates is relatively small.

In a nutshell, note that almost all traditional source compressors are following the hybrid transform coding paradigm to evolve, e.g., mode selection in H.265 and H.266. Accordingly, the idea of signal-dependent transform in traditional source compression codecs inspires us to upgrade plain NeurJSCC to a source-and-channel-dependent adaptive mode. The marriage of explicit NeurJSCC (primary) and implicit NeurJSCC (auxiliary) can contribute to significant performance improvement as presented later.

\subsubsection{\underline{Hybrid NeurJSCC Dominated by Implicit Component}}

Although implicit NeurJSCC has been successful in transmitting continuous signal especially for some emerging data, one of the fundamental challenges is their inability to scale to complex signals due to the limited approximation ability of neural fields. To tackle this, we need a suitable \emph{implicit prior} indicating the signal local feature, which can enhance the signal reconstruction. As an evolved version of implicit NeurJSCC, this class of hybrid NeurJSCC techniques still depend on INR as the primary functionality whilst integrating feature embeddings as conditions. As shown in Fig. \ref{Fig2}, unlike the conventional non-feature aided neural fields that embed coordinates with a fixed neural function such as positional Fourier embeddings, our upgraded neural fields in hybrid NeurJSCC determine the output quantities $\boldsymbol{q}$ using the coordinate $\boldsymbol{x}$ and the \emph{latent feature} variable $\boldsymbol{z}$, i.e., $\boldsymbol{q} = f_{\boldsymbol{\theta}}(\boldsymbol{x},\boldsymbol{z})$. This \emph{conditional neural field} lets us vary the field agilely by varying the latent variable, which releases the signal representation complexity in part away from the implicit neural field function into the explicit latent variable.

The critical latent variables $\boldsymbol{z}$ in hybrid NeurJSCC could be samples from arbitrary distribution or semantic labels indicating the object and scene properties. For example, when we want to transmit a 3D car object via hybrid NeurJSCC, by conditioning the neural field on semantic labels describing the shape and appearance of this car, we can synthesis the car with free view at the receiver. If the latent variables $\boldsymbol{z}$ are defined on a smooth semantic space, they can be interpolated and edited \cite{xie2022neural}. We then discuss useful techniques about how to generate the latent variables $\boldsymbol{z}$.

\begin{itemize}
  \item \emph{Feed-forward encoder for $\boldsymbol{z}$:} An intuitive solution to find the latent variable $\boldsymbol{z}$ is generated via a neural encoder from observations $\boldsymbol{o}$. Parameters of this encoder is jointly optimized with the conditional neural field function over a set of data and tasks. This amortized encoder function is fast to generate the latent code $\boldsymbol{z}$ with only the forward pass operation, which contributes to the fast inference of the whole hybrid NeurJSCC.

  \item \emph{Auto-decoder for $\boldsymbol{z}$:} It works by performing stochastic optimization on the latent code $\boldsymbol{z}$ directly that minimizes the signal reconstruction error at the receiver \cite{park2019deepsdf}. This instance-adaptive optimization acts as encoder to map the observation $\boldsymbol{o}$ to its corresponding latent code $\boldsymbol{z}$. Unlike the feed-forward encoder, auto-decoder does not impose constraints about the observations. For example, a well-learned latent encoder for image observations requires $\boldsymbol{o}$ to be on a pxiel grid with a given distribution, but auto-decoder provides robustness to support observations of arbitrary format and distribution because the gradient of the loss with respect to $\boldsymbol{z}$ can be computed independently of $\boldsymbol{o}$. The longer inference time cost of auto-decoder exchanges a much better generalization.

  \item \emph{Global conditioning:} It specifies the neural field function using a single latent code $\boldsymbol{z}$ across all coordinates. For a better performance, we expect $\boldsymbol{z}$ being as compact as possible, which tries its best to carry more information with a given dimension. However, it is difficult to achieve this vision in practice. A global latent code can succeed at representing distributions of few degrees of freedom (such as the semantic label of a single object), but cannot represent complex scenes containing many independent ingredients.

  \item \emph{Local conditioning:} This method is widely employed in emerging neural field techniques, where the latent code is obtained through a \emph{coordinate-dependent} function $\boldsymbol{z} = g(\boldsymbol{x})$. By this approach, the neural field function becomes coordinate-specific, which can represent complex signals more efficiently. The query function $g$ is often designed as a lookup from multiple parametric embeddings that are referred to as the \emph{feature grids} of multiple levels of detail (LODs). Every feature grid is a sparse discrete data structure storing feature vectors at vertices of grids such as 2D pixel grids, 3D voxel grids, etc. The latent variable $\boldsymbol{z}$ at arbitrary position is queried from its neighborhood explicit feature vectors using interpolation algorithms as shown in Fig. \ref{Fig2}.

\end{itemize}

Following the well-optimized conditional INR functions, we need to design source-channel coding to transmit both implicit model parameters and explicit feature grids. Since neural fields use simple MLP, the transmission budget of model parameters is trivial compared to that of feature grids. Accordingly, We put more efforts on the high-efficiency and robust transmission of feature grids. We follow the idea of auto-decoder by performing stochastic optimization to learn multiple feature grids at LODs directly. During this procedure, we incorporate a \emph{latent prior} on the components of feature grids as the entropy penalty like that in explicit NeurJSCC, which guides the downstream NeurJSCC rate allocation to maximize the system coding gain. Apart from the transmission rate control within each feature grid, our hierarchical representation also supports progressive and variable transmission of feature grids of different LODs, which scales the end-to-end transmission quality smoothly. By this approach, we can achieve a very flexible end-to-end RD trade-off. As shown in Fig. \ref{Fig2}, these neural field parameters, latent feature grids of LODs, and the latent prior model are all learned jointly.

\section{Applications of NeurJSCC}

We start our discussion by a comprehensive comparison among various coded transmission paradigms. We then provide exemplary results with respect to their application domains.

\subsection{Comprehensive Comparisons}

Table \ref{Table1} provides a comprehensive comparison of different coded transmission schemes. We list source media types that each coding paradigm is more expert to deal with, and horizontally compare their characters in terms of coding speed, RD performance, robustness, perceptual quality, scalability, etc. In general, as the fruit of years of research and development, classical coded transmission systems (e.g., VVC \cite{bross2021developments} + 5G LDPC \cite{richardson2018design} for wireless video transmission) have been widely applied for transmitting traditional structured media. However, due to the separated source and channel coding design, it suffers from time-varying channel conditions, in which case the mismatch between communication rate and channel capacity leads to obvious \emph{cliff-effect}, i.e., the performance breaks down when the channel capacity goes below communication rate. Powerful mechanisms (e.g., HARQ) must be employed to ensure no error left in channel decoding, which however leads to higher latency and more resource occupation. In comparison, both explicit and implicit NeurJSCC methods benefit from the integrated design of source and channel components, which provides robustness and better adaptability to the requirements of human perception and machine tasks. Their corresponding hybrid versions further improve performance and flexibility without sacrificing decoding speed, which makes sense for popularizing these techniques.

\begin{table*}[t]
\renewcommand{\arraystretch}{1.5}
  \centering
  \small

  \caption{Comparisons of different coded transmission systems.}

  \begin{tabular}{m{4cm}|m{4cm}|m{4cm}|m{4cm}}

    \Xhline{1pt}

      & \centering \multirow{2}*{{{\makecell{\textbf{Classical coded transmission} \\ \emph{(hand-crafted)}}}}} & \multicolumn{2}{c}{\textbf{NeurJSCC transmission} \emph{(learning-based)}} \tabularnewline

    \cline{3-4}
     &  &  \centering Explicit / Hybrid (explicit) & \centering Implicit / Hybrid (implicit) \tabularnewline

     \Xhline{1pt}

     \centering Suitable source media type & \multicolumn{2}{c|}{Traditional structured media including audio, image, video, etc.} &  Emerging media including Gigapixel image, 8K+ video, 3D shape, realistic 3D scene, etc. \tabularnewline

     \hline
     \centering Encoding speed & \centering {Medium to {\textbf{Fast}}} & \centering Medium / Slow & \centering Very slow / Slow \tabularnewline

     \hline
     \centering Decoding speed & \centering {{Medium}} & \centering Slow & \centering \textbf{Fast} \tabularnewline

     \hline
     \centering End-to-end RD performance & \centering {{Good}} & \centering Very good / \textbf{Excellent} & \centering Fair / Very good \tabularnewline

     \hline
     \centering Robustness to wireless channel & \centering {{Fair (cliff effect)}} & \multicolumn{2}{c}{\textbf{Good} (no cliff effect, graceful degradation with channel SNR)} \tabularnewline

     \hline
     \centering Human perceptual quality & \centering {{Good}} & \centering  Very good / \textbf{Excellent} & \centering \emph{To be explored} \tabularnewline

    \hline
     \centering Scalability to machine tasks  & \centering {{Needs work}} & \centering Adequate / Flexible & \centering \textbf{Very flexible} \tabularnewline

    \Xhline{1pt}
  \end{tabular}
  \label{Table1}
\end{table*}

\begin{figure*}
\centering{\includegraphics[scale=0.8]{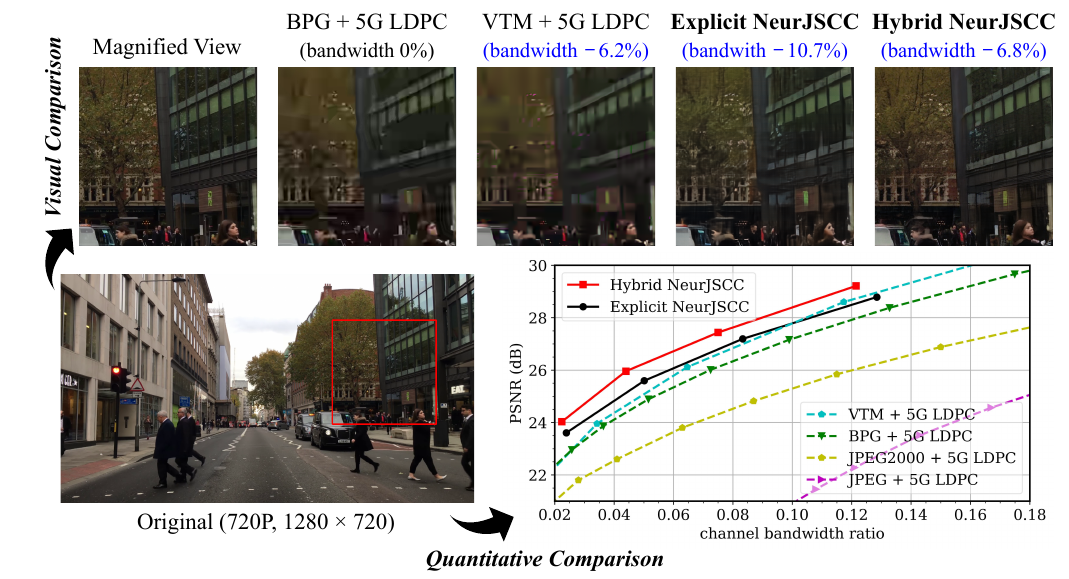}}
	\caption{Transmission RD results and visual comparisons of different coded transmission systems, where blue number indicate the percentage of bandwidth cost saving compared to the ``BPG + 5G LDPC'' scheme whose channel bandwidth ratio (denoting the ratio of the channel-input dimension to the source data dimension) is 0.0355. Our explicit/hybrid NeurJSCC schemes provides visually appealing reconstructions with the lower wireless channel bandwidth cost.}\label{Fig3}
	\vspace{-1em}
\end{figure*}

\begin{figure*}
\centering{\includegraphics[scale=0.5]{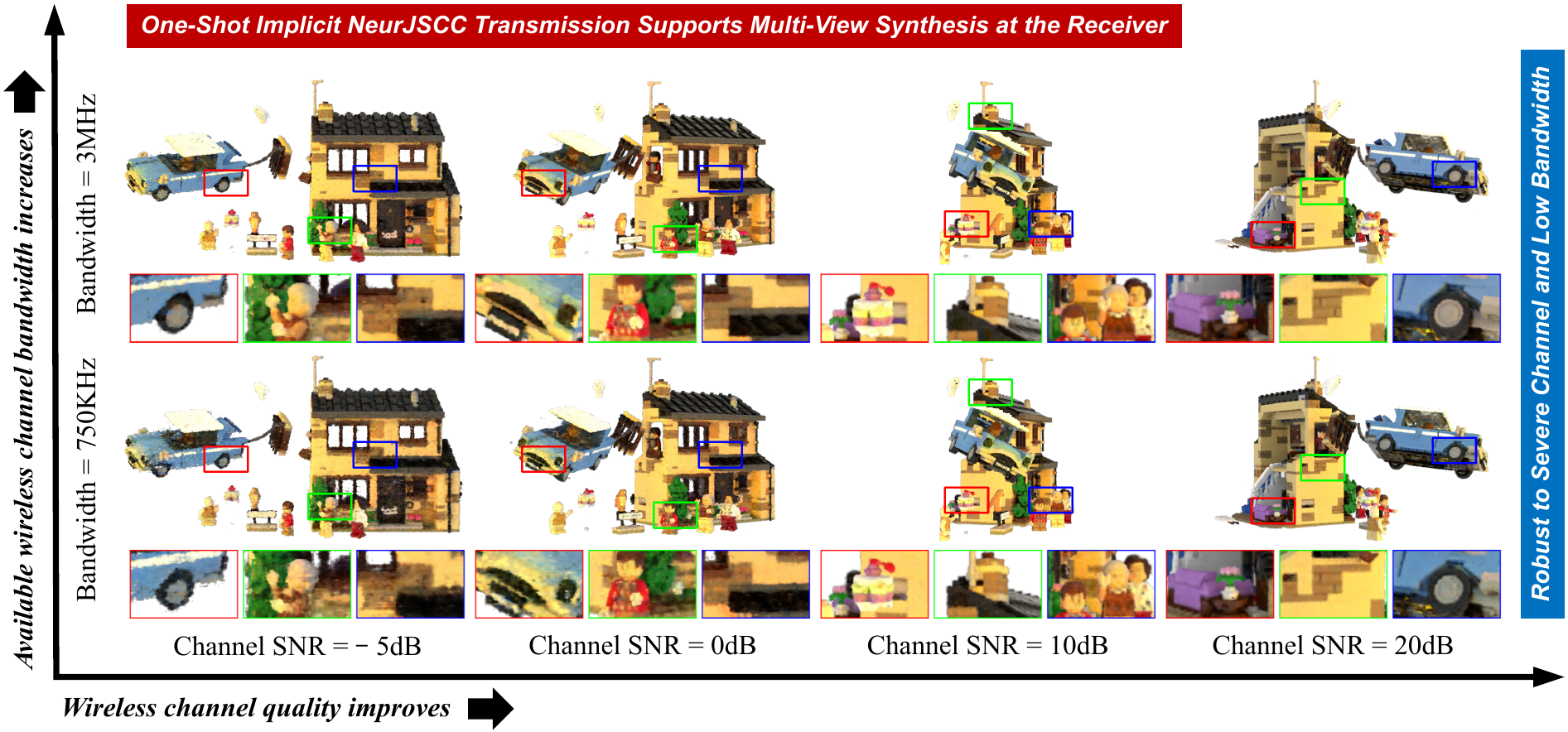}}
	\caption{Realistic 3D scenes can be transmitted over wireless channels using our implicit/hybrid NeurJSCC with high-fidelity details. For better presentation, three colored rectangles are shown to highlight details. In this article, we only provide the reconstruction from a single viewpoint for each case, full rendering results at any view are available at the website \emph{https://semcomm.github.io/semcoding}.}\label{Fig4}
	\vspace{-1em}
\end{figure*}

\subsection{Explicit/Hybrid NeurJSCC for Image Transmission}

As one exemplary application, we apply explicit NeurJSCC and its hybrid version for end-to-end image transmission. They are compared with many classical coded transmission schemes including  the state-of-the-art (SOTA) engineered image codec VTM (intra-frame coding scheme of the VVC (H.266) standard) \cite{bross2021developments}. The experiments are taken over the AWGN channel at signal-to-noise ratio (SNR) 0dB (severe channel condition). Results are provided in Fig. \ref{Fig3}, our NeurJSCC presents superior end-to-end RD performance over classical coded transmission schemes. Besides, NeurJSCC methods synthesize convincing details that are appealing for human vision, but the comparison schemes tend to produce blocky/blurry blobs for regions due to insufficient bandwidth.

\subsection{Implicit/Hybrid NeurJSCC for 3D Scene Transmission}

To intuitively present the advantages of implicit NeurJSCC and its hybrid version, we apply them for end-to-end transmission of a 3D scene composed of LEGO toys as shown in Fig. \ref{Fig4}.

Classical systems transmit a 3D scene in two steps. It first transmits each of the original views using traditional multi-view coded transmission protocol (e.g., VVC + 5G LDPC), then performs rendering at the receiver. Limited by the narrow range of each viewpoint, it requires hundreds of megabytes of bandwidth to transmit a prohibitively large number of original views to synthesis arbitrary-viewpoint 3D scenes. This naive solution prohibits XR applied to mobile devices for exploring emerging visual media.

As an alternative, our NeurJSCC approaches integrate both transmitting and rendering steps by directly fitting the 3D scene over various wireless channel conditions. Fig. \ref{Fig4} presents the reconstructed renderings in under different channel SNRs and bandwidth constraints. Here, the channel bandwidth cost is computed within 1s according to 5G OFDM configurations: 14 OFDM symbols@1ms with a subcarrier bandwidth 15KHz. Our NeurJSCC transmission scheme can produce realistic renderings over wireless channels, even under severe channel conditions and meager bandwidth budgets. In addition, the rendering quality improves gracefully with the better wireless channel conditions and the increases in bandwidth cost. That verifies the robustness of implicit/hybrid NeurJSCC.

\section{Conclusion and Forward Looking}

As we have seen, NeurJSCC has rich dimensions evolving toward different technical routes, which are suitable for transmitting different types of source media. In our view, there are several factors driving this rapid progress.

First, explicit NeurJSCC and its hybrid version boost the end-to-end transmission performance by leveraging the nonlinear and overfitting properties of neural networks. For traditional structured data, techniques such as nonlinear transform, variational latent prior modeling, variable-length NeurJSCC, and joint learning with wireless channels have improved end-to-end RD performance, leading to clear leaps over classical coded transmission schemes. Second, implicit NeurJSCC and its hybrid version provide a fundamentally different paradigm for complex signal transmission. They leverage INR functions to parameterize continuous fields using simple MLP without the need of complex neural networks, which reduce the neural network entry barrier and processing complexity especially for emerging unstructured data. Following that, NeurJSCC focuses on the transmission of INR model parameters and explicit feature grids. Applications in wireless 3D scene transmission supporting free view synthesis in XR will further popularize this NeurJSCC paradigm.

Despite this progress, we believe that NeurJSCC for end-to-end communication have only started to scratch the surface and there still remains great potential to be explored. In terms of \emph{techniques}, a common limit of explicit NeurJSCC related methods is their generalization ability to unseen data and tasks using the amortized model. Our adaptive coding mechanism in hybrid NeurJSCC can somewhat mitigate this through additional computation overhead in online updating, but we believe that integrating stronger prior can further enable explicit NeurJSCC related methods to generalize and perform better. Other task specific knowledge can be useful. Besides, a common limit of implicit NeurJSCC related methods is their slow encoding speed due to the use of online stochastic optimization. We believe integration of stronger meta-learning methods can contribute to faster encoding. Other inductive biases such as law of physics and network architectures can further help speeding up and better performance.

In terms of \emph{applications}, most NeurJSCC approaches focus on the ``low-level'' transmission tasks (e.g., data or scene reconstruction at the receiver). Their extension to ``high-level'' semantic tasks including media understanding, 3D scene interaction, and other machine tasks still requires extensive research. In that case, the semantic importance modeling on the latent feature space will include not only the variationally estimated entropy but also other task related metrics, which is a valuable research direction. Furthermore, current NeurJSCC focuses on single data modality, novel NeurJSCC techniques supporting \emph{multiple modalities} might be a fruitful research topic. For example, implicit NeurJSCC related methods synthesis fields based on multiple modalities input can boost the end-to-end transmission performance.

\ifCLASSOPTIONcaptionsoff
  \newpage
\fi

\bibliographystyle{IEEEtran}

\bibliography{Ref}

\section*{Biographies}

\small{

\emph{Sixian Wang} is currently at Beijing University of Posts and Telecommunications (BUPT). His research interests include source and channel coding, visual computing.

\emph{Jincheng Dai} is currently an associate professor at BUPT. His research interests include neural compression, neural transmission, and semantic communications.

\emph{Xiaoqi Qin} is currently an associate professor at BUPT. Her research interests include distributed learning, age of information.

\emph{Kai Niu} is currently a professor at BUPT. His research interests include information theory, polar codes.

\emph{Ping Zhang} is currently a professor at BUPT. He is a Fellow of IEEE and an Academician of Chinese Academy of Engineering. His  research interests include wireless communications.

}

\end{document}